\documentclass[journal=jacsat,manuscript=article]{achemso}

\usepackage[version=3]{mhchem} 
\usepackage{color}
\usepackage{wasysym}

\newcommand{\inlinecite}[1]{Ref.~\citenum{#1}}

\author{Ruiwen Xie}
\altaffiliation{Equal Contribution}
\email{ruiwen.xie@tmm.tu-darmstadt.de}
\affiliation[TU Darmstadt]
{Theory of Magnetic Materials Group, Department of Materials and Geosciences, Technical University of Darmstadt, Otto-Berndt-Str. 3, 64287 Darmstadt, Germany}
\author{Maximilian Mellin}
\affiliation{Surface Science Laboratory, Department of Materials and Geosciences, Technical University of Darmstadt, Peter-Grünberg-Str. 4, 64287 Darmstadt, Germany}
\altaffiliation{Equal Contribution}
\author{Wolfram Jaegermann}
\affiliation{Surface Science Laboratory, Department of Materials and Geosciences, Technical University of Darmstadt, Peter-Grünberg-Str. 4, 64287 Darmstadt, Germany}
\author{Jan P. Hofmann}
\email{hofmann@surface.tu-darmstadt.de}
\affiliation{Surface Science Laboratory, Department of Materials and Geosciences, Technical University of Darmstadt, Peter-Grünberg-Str. 4, 64287 Darmstadt, Germany}
\author{Frank M. F. de Groot}
\affiliation{Debye Institute for Nanomaterials Science, Utrecht University, 3584 CG Utrecht, Netherlands}
\author{Hongbin Zhang}
\affiliation
{Theory of Magnetic Materials Group, Department of Materials and Geosciences, Technical University of Darmstadt, Otto-Berndt-Str. 3, 64287 Darmstadt, Germany}

\title[An \textsf{achemso} demo]
  {Redox Chemistry of LiCoO$_2$, LiNiO$_2$, and LiNi$_{1/3}$Mn$_{1/3}$Co$_{1/3}$O$_2$ Cathodes: Deduced via XPS, DFT+DMFT, and Charge Transfer Multiplet Simulations
}

\abbreviations{IR,NMR,UV}

\begin{document}







\begin{abstract}
Understanding the evolution of the physicochemical bulk properties during the Li deintercalation (charging) process is critical for optimizing battery cathode materials. In this study, we combine X-ray photoelectron spectroscopy (XPS), density functional theory plus dynamical mean-field theory (DFT+DMFT) calculations, and charge transfer multiplet (CTM) model simulations to investigate how hybridization between transition metal (TM) 3$d$ and oxygen 2$p$ orbitals evolves with Li deintercalation. 
Based on the presented approach combining theoretical calculations and experimental studies of pristine and deintercalated cathodes, two important problems of ion batteries can be addressed: i) the detailed electronic structure and involved changes with deintercalation providing information of the charge compensation mechanism, and ii) the precise experimental analysis of XPS data which are dominated by charge transfer coupled to final-state effects affecting the satellite structure.
As main result for the investigated Li TM oxides, it can be concluded that the electron transfer coupled to the Li$^{+}$-ion migration does not follow a rigid band model but is modified due to changes in TM 3$d$ and O 2$p$ states hybridization. 
Furthermore, this integrated approach identifies the 2$p$ XPS satellite peak intensity of TM as an effective indicator of the redox chemistry. With that the redox chemistry of cathodes can be deduced, thus offering a foundation for designing more efficient battery materials.
\end{abstract}

\section{Introduction}
The performance of intercalation battery materials depends on a number of interrelated effects induced by changes of structure \cite{reimers1992electrochemical,wang2004electrochemical} and chemical composition, as a consequence of the exchange of Li$^{+}$ ions and electrons, which modify the bulk as well as the surface properties. \cite{gauthier2018probing,schulz2018xps,hausbrand2020electronic,hausbrand2017surface} 
Especially, the changes in the electronic structure of transition metal oxide cathode materials during deintercalation are complex, involving changes in oxidation states, insulator-metal transitions affecting the electronic conductivity, non-rigid band behaviour in the valence and conduction bands, lowering of the Fermi level, oxygen involvement in the charge compensation (anionic redox) and oxygen release.\cite{ensling2014nonrigid, naylor2019depth, flores2021operando,uchimoto2001changes, jung2017oxygen}     
The utilization of anionic redox has gained renewed attention in Li-rich cathode materials (e.g., $\mathrm{Li}_{1} \mathrm{Mn} \mathrm{O}_3$, $\mathrm{Li}_{1.2} \mathrm{Mn}_{0.6} \mathrm{Ni}_{0.2} \mathrm{O}_2$), due to its ability to provide extra capacity.\cite{li2022improving} 
However, the oxidation of the O$^{2-}$ ions and the related release of oxygen are of high concern due to involvement in degradation \cite{jung2017oxygen} and thermal runaway effects.\cite{li2021thermal}

To explain the interplay between oxygen release and stable charge compensation, Assat and Tarascon proposed a model where the Coulomb interaction of the transition metal (TM) 3$d$ electrons and the charge transfer energy are the key factors.\cite{assat2018fundamental}
They identified three distinct regimes: (1) when the Coulomb interaction is much smaller than the charge transfer energy, charge compensation primarily occurs through cationic redox; (2) when the Coulomb interaction substantially exceeds the charge transfer energy, irreversible anionic redox dominates, often accompanied by the release of oxygen; (3) when the magnitude of charge transfer energy is approximately half of the Coulomb interaction, both cationic and reversible anionic redox coexist (two-band redox).  
Later an advanced framework for classifying reversibility in anion-rich materials was proposed by Yahia et al. by showing that the number of holes per oxygen is critical for sustaining reversible anionic capacity.\cite{ben2019unified} 
In addition, the reversible participation of oxygen in the charge compensation has also been associated with the formation of confined $\mathrm{O}_2$ dimers in the crystal, evidenced by O K-edge RIXS signals.\cite{house2021role} However, similar spectral signatures have also been observed in non-Li-rich cathodes.\cite{menon2023oxygen} 
For instance, the conventional cathode $\mathrm{LiCoO}_2$ (LCO), is classified as a Mott–Hubbard insulator, where the electrochemical behaviour is dominated by cationic redox processes.\cite{van1991electronic}
Nevertheless, the formation of $\mathrm{O}^{1-}$ states was detected based on oxygen K-edge X-ray absorption near-edge structure (XANES) for highly deintercalated LCO thin films, and for the Li-rich NiCoMn-based cathodes at high voltage.\cite{ensling2014nonrigid, liang2025novel} 
In this regard, for LCO the participation of oxygen in the charge compensation could be due to a transition from a Mott-Hubbard to a charge-transfer-dominated system in the deintercalation process.
Additionally, the holes compensating Li$^{+}$ were mainly found in O $2p$ states in Li$_x$Ni$_{1-x}$O for $0.05<x<0.5$.\cite{kuiper1989character} Moreover, oxygen-driven redox mechanisms have been proposed for Ni-rich Li-ion cathode materials. In these systems, previous DMFT calculations suggest a constant charge state of Ni$^{2+}$, while the charge state of oxygen varies from -1.5 in LiNiO$_2$ (LNO) to -1 in NiO$_2$.\cite{genreith2023oxygen}
This highlights the broader relevance of oxygen participation in various cathode materials, which depends on the electronic valence band structure of the intercalated TM oxide and inherent changes during Li (de)intercalation.
Consequently, a refined model is required to gain a dynamic perspective on redox processes, with an emphasis on the evolution of the electronic structure.
In particular, the role of hybridization between TM 3$d$ and oxygen 2$p$ states may be of relevance during delithiation, which can be associated with a non-ridig vs. rigid band behaviour in the involved redox reactions. 

X-ray photoelectron spectroscopy (XPS) provides an efficient method for examining the electronic structure of materials. 
In XPS, a core-hole is created by the emission of the core electron to the vacuum, leading to pronounced final state charge transfer (CT) effects in TM core spectra in systems with open $d$-shell configuration. This is not the case for X-ray absorption spectroscopy (XAS), in which the core-hole is directly screened by the electron transferred into the conduction band. \cite{de2005multiplet}
The final state CT effects manifest themselves in a satellite structure in the 2$p$ XPS spectra of TM, with its intensity being reduced in the case of the deintercalation of LCO.\cite{daheron2008electron}  
Particularly, when combined with charge transfer multiplet (CTM) simulations, XPS enables the retrieval of key parameters by comparing TM core spectra with theoretical models.\cite{zhang2023autonomous} 
Previous theoretical work employing the DFT+DMFT approach on Li$_x$CoO$_2$ has focused on its thermodynamic stability.\cite{isaacs2020compositional}
More recently, a comparative study on the CT effect in LCO and deintercalated LCO was conducted using hard X-ray photoelectron spectroscopy (HAXPES) and CTM simulations,\cite{fantin2023self} in which the spectra of deintercalated LCO were reproduced with a negative CT energy. However, the changes of satellite feature during Li deintercalation was not much discussed. 

The urgent demands for higher practical capacity lead to the creation of mixed TM cathode materials (LiNi$_{1-x-y}$Mn$_x$Co$_y$O$_2$) and especially the Ni-rich compounds.
To the best of our knowledge, the electronic structure of these cathode materials has not been extensively studied, particularly in relation to its core-level spectral features and in comparison to the behaviours of LCO and LNO. 
In this work, we unveil the detailed physical mechanisms behind these spectral features, focusing on the role of electronic structure evolution and its interplay with Li concentration (open circuit-voltage at different charging states) in LiNi$_{1/3}$Mn$_{1/3}$Co$_{1/3}$O$_2$ (LNMCO). By integrating experimental observations with advanced DFT+DMFT and CTM simulations, we provide further understanding of the redox behaviour and the underlying mechanisms of the satellite intensity variations in these complex systems.

\section{Results and discussion}
\subsection{Theoretical electronic structure}

\begin{figure}[h!]
	\centering
	\includegraphics[width=1.0\textwidth]{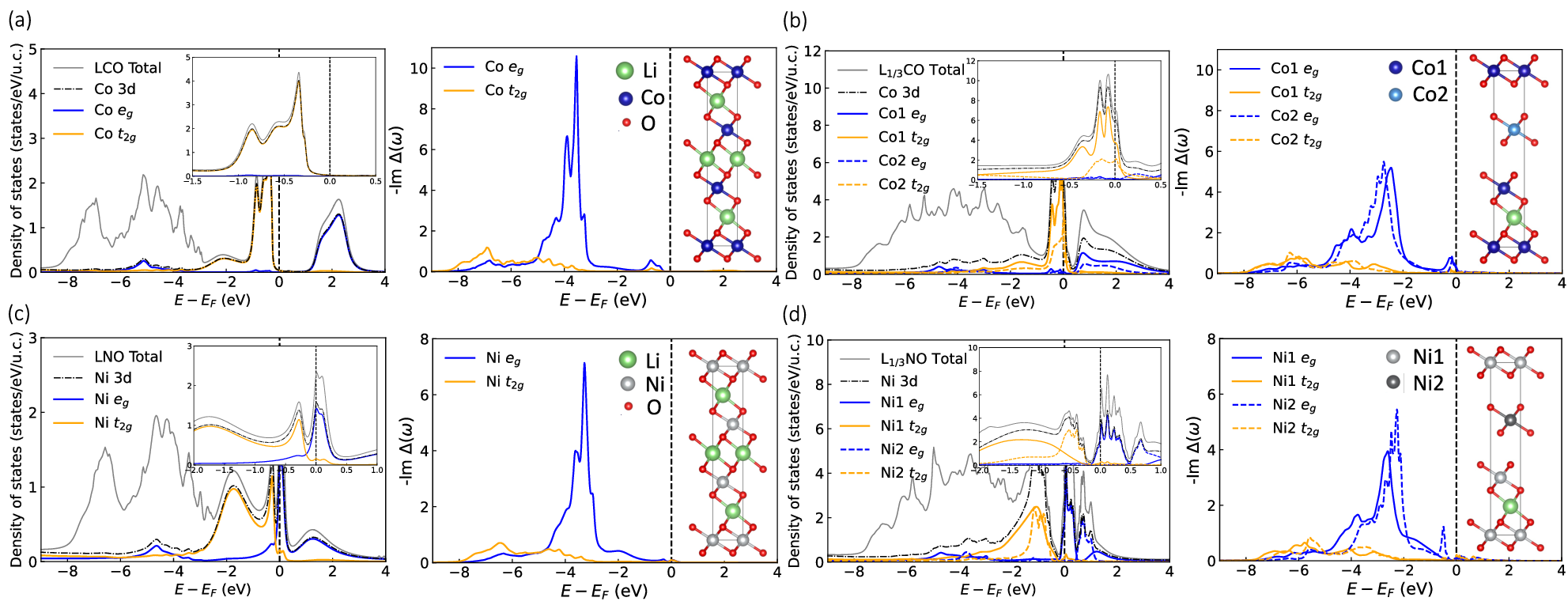}
	\caption{Orbital-resolved density of states (DOS) (left panel) and hybridization function (right panel) of (a) LiCoO$_2$ (LCO), (b) Li$_{1/3}$CoO$_2$ (L$_{1/3}$CO), (c) LiNiO$_2$ (LNO), and (d) Li$_{1/3}$NiO$_2$ (L$_{1/3}$NO). The DOSs close to the Fermi level are zoomed in for more details. The computational cells adopted are shown in the inset of the hybridization function.}
	\label{fig:DMFT_Co}
\end{figure}

We performed charge self-consistent DFT+DMFT calculations to obtain the electronic structures of Li$_x$CoO$_2$ and Li$_x$NiO$_2$ ($x=1$ and 1/3, see Fig.~\ref{fig:DMFT_Co}), in order to study the variations of the density of states (DOS) and the hybridization function during delithiation.
In DMFT, the imaginary part of the hybridization function ($-\mathrm{Im} \Delta(\omega)$) is the energy-resolved covalency: it measures how strongly the TM 3$d$ orbital hybridizes with the O 2$p$ states. A large $-\mathrm{Im} \Delta(\omega)$ indicates strong mixing between 3$d$ and O 2$p$ states. Practically, peaks in $-\mathrm{Im} \Delta(\omega)$ track the O 2$p$ band edges, and $-\mathrm{Im} \Delta(\omega)_{e_g}$ is typically lager than $-\mathrm{Im} \Delta(\omega)_{t_{2g}}$ (see the right panels of Fig.~\ref{fig:DMFT_Co}(a)-(d)), because $\sigma$ bonding to O-2$p$ is stronger than $\pi$.


In this work we focus on the effect of chemical composition by fixing the lattice parameters and the atomic coordinates of Li$_x$CoO$_2$ and Li$_x$NiO$_2$ to those of LCO ($a=2.818$ \AA, $c=14.066$ \AA) and LNO ($a=2.883$ \AA, $c=14.199$ \AA), respectively, while simply removing the Li atom. 
The study of structures with Li concentration $x=1$ and 1/3 involves different types of Co/Ni depending on their local chemical environments, labelled as Co with Li located both below and above the Co/Ni layer, Co1/Ni1 and Co2/Ni2 with one and both neighbouring Li atoms removed, respectively. 


As can be seen from Fig.~\ref{fig:DMFT_Co} (a) and (c), LCO exhibits an insulating character while the band gap at the Fermi energy ($E_F$) is missing for LNO.
For LNO, the Ni 3$d$ orbitals contribute only around 60\% to the total DOS at $E_F$, while the rest originates mainly from oxygen. 
Additionally, at $E_F$ the $e_{g}$ states of Ni in LNO dominate while for Co in LCO, its valence band mainly consists of $t_{2g}$ states, which are visible in the experimentally measured valence band spectra of LCO and LNO (see Fig. S1).
Moreover, LCO converts to a metallic state with Li deintercalation, as can be seen by the lowering of $E_F$ into the $t_{2g}$ orbitals (see Fig.~\ref{fig:DMFT_Co}(b)).\cite{nishizawa1998irreversible}
On the one hand, the energy level corresponding to $E_F$ moves downwards by removing positive Li ions and the hybridization peaks are shifted towards $E_F$ during delithiation (by comparing the right panels of Fig.~\ref{fig:DMFT_Co}(a) and (b), or (c) and (d)); on the other hand, the maximal peak intensity of the hybridization function is reduced, which indicates a transition from localized to more itinerant behaviour. 
By comparing Fig.~\ref{fig:DMFT_Co}(b) and (d), an intriguing difference between Co and Ni with Li deintercalation is noticed. The hybridization peak of Co2 is slightly shifted downwards with reference to $E_F$ as compared to that of Co1, whereas for Ni the trend is reversed. 
This phenomenon, together with the existing oxygen DOS at $E_F$ in LNO, suggests that more mixed $3d$-$2p$ character at finite binding energies (near-$E_F$ region) is expected in Li$_x$NiO$_2$ before and after delithiation, which could explain the more largely involved anionic redox in the cyclic process of Li$_x$NiO$_2$.

\begin{figure}[h!]
 \centering
  \includegraphics[width=1.0\textwidth]{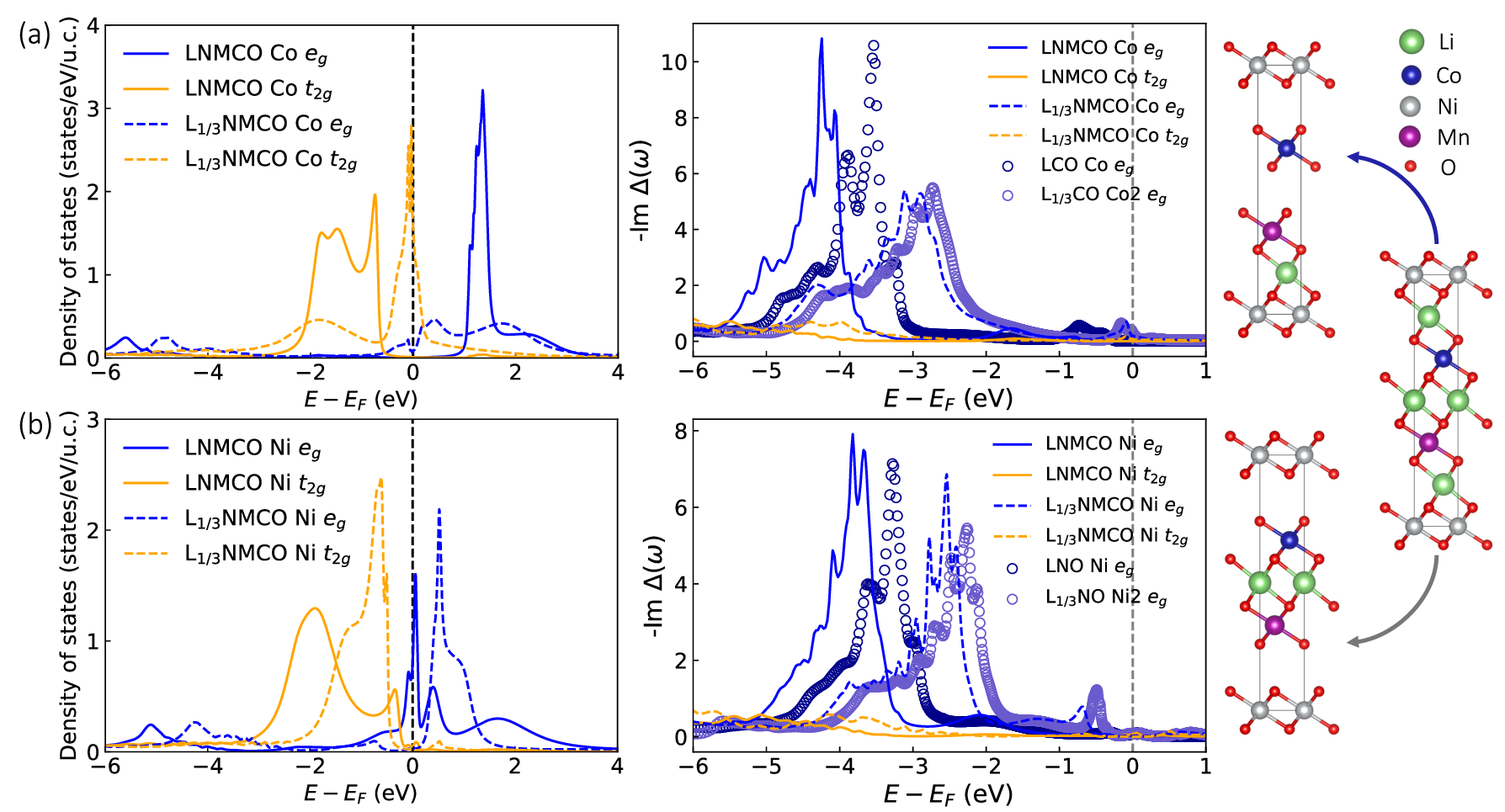}
  \caption{Orbital-resolved density of states (DOS) (left panel) and hybridization function (right panel) of (a) Co in LiNi$_{1/3}$Mn$_{1/3}$Co$_{1/3}$O$_2$ (LNMCO) and Li$_{1/3}$Ni$_{1/3}$Mn$_{1/3}$Co$_{1/3}$O$_2$ (L$^{\mathrm{Co}}_{1/3}$NMCO). The hybridization functions of Co in LiCoO$_2$ (LCO) and Co2 in Li$_{1/3}$CoO$_2$ (L$_{1/3}$CO) are shown for comparison, and (b) Ni in LNMCO and L$_{1/3}$NMCO (L$^{\mathrm{Ni}}_{1/3}$NMCO). The hybridization functions of Ni in LiNiO$_2$ (LNO) and Ni2 in Li$_{1/3}$NiO$_2$ (L$_{1/3}$NO) are shown for comparison.
  The simulation cells corresponding to LNMCO, L$^{\mathrm{Co}}_{1/3}$NMCO, and L$^{\mathrm{Ni}}_{1/3}$NMCO are illustrated on the right side.}
  \label{fig:nmc}
\end{figure}

In contrast to LNO, LNMCO shows good stability with charging/discharging cycle. The low lying bands of Mn $t_{2g}$ states related to Mn$^{4+}$ are not redox active, thus serving as a structural stabilizer.\cite{cherkashinin2015electron,manthiram2020reflection} 
In order to study the impact of the TM mixing on the electronic structure, we further performed the DFT+DMFT calculations on Li$_x$Ni$_{1/3}$Mn$_{1/3}$Co$_{1/3}$O$_2$ (x = 1, 1/3) adopting a simplified crystal structure as displayed in Fig.~\ref{fig:nmc}. 
It can be seen from Fig.~\ref{fig:nmc}(a) that with Li deintercalation, the energy gap between Co $t_{2g}$ orbitals in the valence band and $e_g$ orbitals in the conduction band disappears, while it tends to open up for Ni (see Fig.~\ref{fig:nmc}(b)).
Additionally, in LNMCO, the hybridization peak position of Ni is slightly closer to $E_F$ as compared to that of Co (see right panels of Fig.~\ref{fig:nmc}(a) and (b)), which is consistent with the picture demonstrated by comparing Ni in LNO and Co in LCO (see Fig.~\ref{fig:DMFT_Co}). And similarly, the hybridization peaks are shifted towards $E_F$ with Li deintercaltion. 
Nevertheless, the presence of Mn drives the hybridization peaks of both Co and Ni farther away from $E_F$ in both pristine and deintercalated structures, which well demonstrates the role of Mn in redistributing the spectral weight with suppressed charge fluctuations caused by ligand-metal charge transfer. 


\begin{figure}[h!]
	\centering
	\includegraphics[width=1\textwidth]{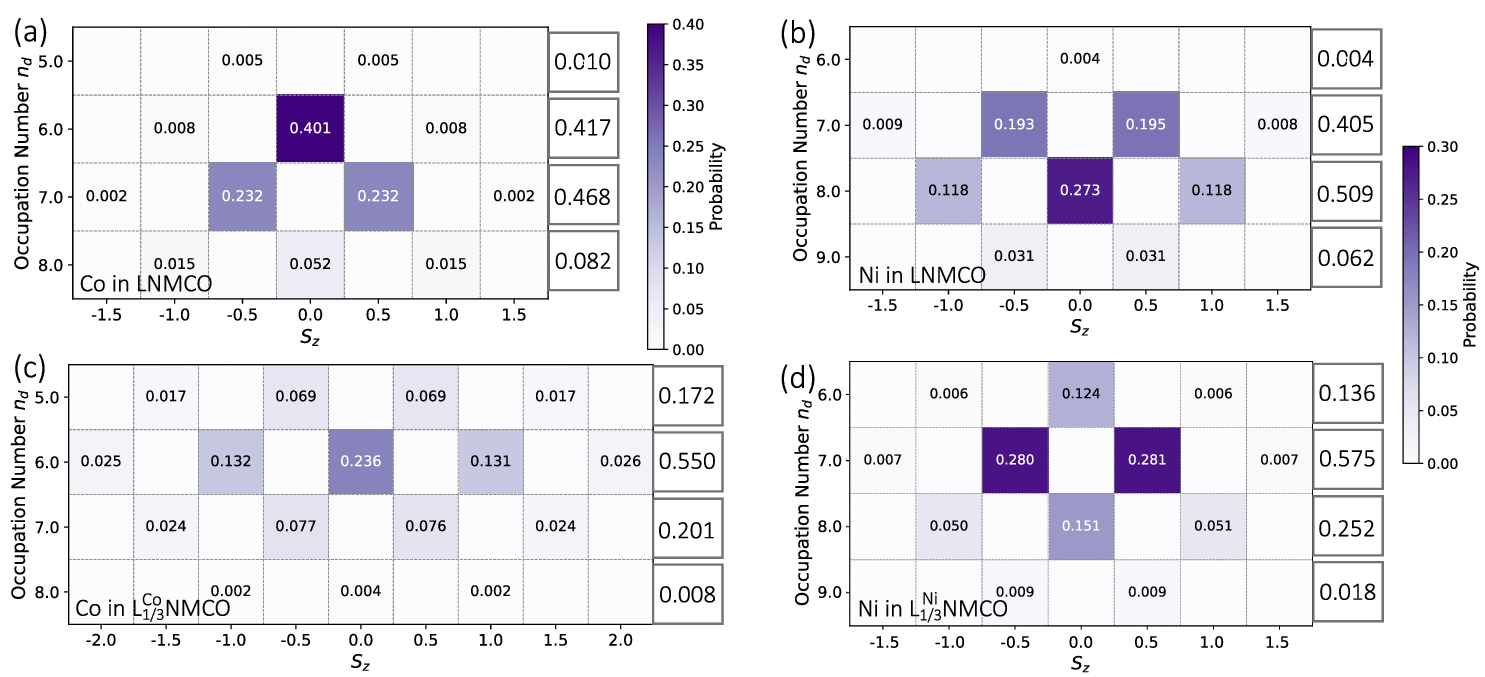}
	\caption{The occupation probabilities as a function of occupation number $n_d$ and expectation value of $S_z$ ($z$-component of the spin angular momentum operator) given by the CTQMC impurity solver in DMFT calculations for (a) Co 3$d$ shell in LNMCO, (b) Ni 3$d$ shell in LNMCO, (c) Co 3$d$ shell in L$^{\mathrm{Co}}_{1/3}$NMCO, and (d) Ni 3$d$ shell in L$^{\mathrm{Ni}}_{1/3}$NMCO. The total probability corresponding to each $n_d$ is given in the box on the right side of each figure. Note that the total probability does not add to 1 because we omit those configurations with probability smaller than 0.0005.}
	\label{fig:nmc_occu}
\end{figure}

We further examine the configurational distributions of Co and Ni 3$d$ shells in LNMCO and L$_{1/3}$NMCO.
It should be noted that the calculated occupation probability depends on the adopted double counting energy. Nevertheless, the trend of configurational probability with respect to Li concentration remains intact. 
Due to strong hybridization with the surrounding ligands, the TM site can exhibit a mixed-valence character.
As demonstrated in Fig.~\ref{fig:nmc_occu}, in LNMCO, the Co$^{2+}$ ($n_d=7$) and Co$^{3+}$ ($n_d=6$), Ni$^{2+}$ ($n_d=8$) and Ni$^{3+}$ ($n_d=7$) states are much more prominent than the Co$^{4+}$ ($n_d=5$) and Ni$^{4+}$ ($n_d=6$) states, respectively. With Li deintercalation, the Co$^{3+}$/Ni$^{3+}$ and Co$^{4+}$/Ni$^{4+}$ states further get enriched while the Co$^{2+}$/Ni$^{2+}$ states get diminished. Therefore, on average the oxidation states of Co and Ni in Li$_x$Ni$_{1/3}$Mn$_{1/3}$Co$_{1/3}$O$_2$ would increase with Li deintercalation.

\begin{figure}[h!]
	\centering
	\includegraphics[width=1.0\textwidth]{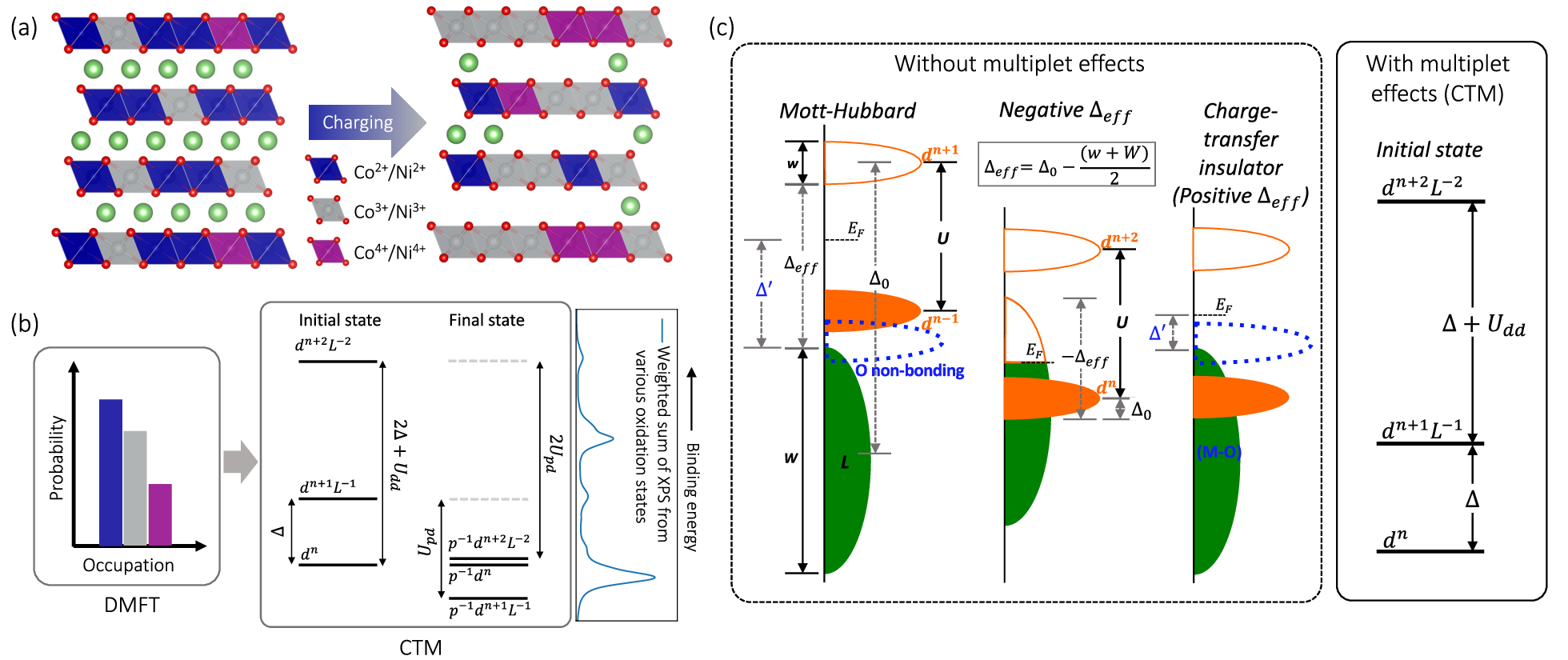}
	\caption{Schematic representations of (a) charge compensation mechanism in Li-ion battery during charging process; (b) the idea of utilizing DMFT-predicted TM $d$-state occupation probabilities corresponding to different cathode oxidation states to simulate the TM 2$p$ XPS based on charge transfer multiplet (CTM) model; and (c) different definitions of Coulomb interaction and charge transfer energy in the electron removal and addition spectra (adopted from \inlinecite{assat2018fundamental, bisogni2016ground, pavarini2016quantum}), and in the CTM model.}
	\label{fig:schema}
\end{figure}

In our previous work, we demonstrated that the satellite intensities in the Co 2$p$ and Ni 2$p$ XPS core spectra of LNMCO decrease with Li deintercalation, with the most pronounced changes occurring at the applied deintercalation voltages corresponding to the formal changes of Co$^{3+}$/Co$^{4+}$ and Ni$^{2+}$/Ni$^{3+}$/Ni$^{4+}$ oxidation states.\cite{mellin2024voltage}
In Fig.~\ref{fig:schema}(a) and (b) we show schematically the changes of state occupation with deintercalation and how these will affect the XPS results considering different final state occupations. 
In this work, we have quantitatively analyzed the contribution of each formal oxidation state using the DFT+DMFT method. Subsequently, we derived the change of TM 2$p$ XPS during delithiation using an ab-initio approach. 
The main idea is to use DMFT results as a guidance for the subsequent CTM simulations of TM 2$p$ XPS, and further to infer the redox activity, as well as the cathode performance. In previous work,\cite{assat2018fundamental, bisogni2016ground, pavarini2016quantum}  the relevant properties, e.g, the voltage and capacity, of cathode materials were usually related to the material-specific parameter including Coulomb interaction and charge transfer energy as derived from the schematic plot of the electronic structure (see Fig.~\ref{fig:schema}(c)). 
Here, we would like to draw the readers' attention to the different definitions of \textit{charge transfer energy} in the schematic electron removal and addition spectra, and in the CTM model,\cite{zaanen1985band} for a consistent understanding. 
For high-oxidation-state compounds, $\Delta_{eff}$ ($\Delta_0-\dfrac{w+W}{2}$, $w$ and $W$ being the bandwidths of upper Hubbard band (UHB) and ligand band, respectively), which is the energy separation between the bottom of UHB and the top of ligand band, is a more relevant definition of charge transfer energy than $\Delta_0$, which is simply the energy difference between the center of the UHB and the ligand band. 
In a different manner, Assat and Tarascon defined the charge transfer energy as $\Delta^\prime$, which is the energy difference between the metal-oxygen (M-O) bonding states and the Fermi level ($E_F$).\cite{assat2018fundamental}
However, these definitions neglect the multiplet effects, i.e., higher multipole Coulomb interactions. When multiplets are included, the charge transfer energy ($\Delta$) is defined as the energy difference between the lowest $d^n$ and $d^{n+1}L^{-1}$ states in the CTM model. 
Further details of the CTM model, and the simulated TM 2$p$ XPS, will be provided in a later section. In the next section, we first discuss the experimentally measured 2$p$ core-level photoemission spectra, which later will be used for comparison with respect to the simulated spectra.

\subsection{Photoelectron spectroscopy}
\begin{figure}[ht!]
 \centering
  \includegraphics[width=0.7\textwidth]{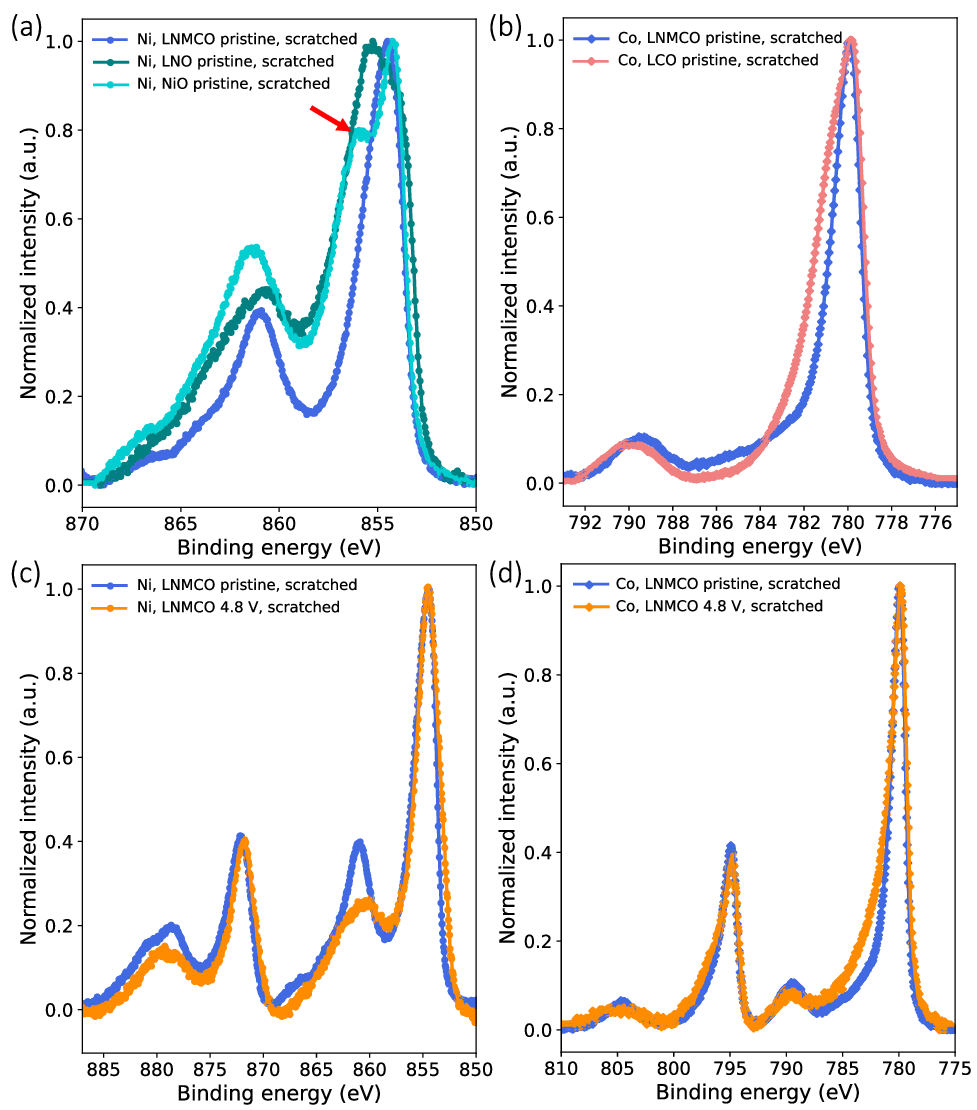}
  \caption{Background subtracted and energy calibrated core spectra  of (a)  Ni 2$p$ XPS of NiO, LNO and LNMCO (b) Co 2$p$ XPS of in-vacuo scratched pristine LCO and LNMCO indicating the non-local screening effects. (c) Ni 2$p$ XPS and (d) Co 2$p$ XPS of in-vacuo scratched pristine LNMCO as compared to the \textit{in-vacuo} scratched fully deintercalated (charged to 4.8 V) LNMCO.}
  \label{fig:fig1}
\end{figure}

In this section, we examine the spectral features of LNMCO in comparison to LCO, LNO, and NiO. As illustrated in Fig.~\ref{fig:fig1}(a), the Ni 2$p$ XPS of NiO shows the characteristic splitting of the main peak (highlighted by the red arrow), due to the non-local screening effect, involving Ni$^{2+}$ states in the neighbouring positions.\cite{van1993nonlocal} 
Both NiO and LNO exhibit a similarly broadened main peak, in contrast to the narrower peak observed in LNMCO. This difference suggests that non-local screening is much less significant in LNMCO, likely due to changes in the occupation of the neighbouring positions by different elements.~\cite{altieri2000core} Consequently, LNMCO is a more suitable system for analyzing the electronic structure evolution during Li deintercalation, as CTM simulations can be performed without additional complications arising from the non-local screening effects.\cite{green2016bond}
A similar trend is observed, albeit to a lesser extent, when comparing the Co 2$p$ photoemission spectra of LCO and LNMCO, where the main peak of LNMCO appears less broadened (see Fig.~\ref{fig:fig1} (b)). 

In Fig.~\ref{fig:fig1}(c) and (d), the spectral features of the Ni 2$p$ and Co 2$p$ photoemission spectra exhibit significant changes with Li deintercalation.
Especially in the Ni 2$p$ XPS, the intensities of the satellites (around 861 eV and 880 eV) decrease for both the 2$p_{3/2}$ and 2$p_{1/2}$ emission peaks, respectively, while their binding energy positions remain approximately unchanged.
With delithiation, the reduction in the intensities of the main Co 2$p$ XPS satellites is much less prominent. 
The overall XPS features of Co remain quite similar in LCO (see Fig. S2) and LNMCO.\cite{daheron2009surface,mellin2024voltage}
Additionally, as observed in Fig.~\ref{fig:fig1}(d), the left shoulder of the main peak slightly increases, which may be directly assigned to an increase of the Co$^{4+}$ oxidation state and a higher binding energy,\cite{mellin2024voltage} or to the non-local screening effects with structural changes.\cite{ensling2014nonrigid,fantin2023self} 
These spectral features will be examined in connection with the changes in the configurational probabilities with delithiation as listed in Fig.~\ref{fig:nmc_occu}, and the following CTM simulations.  

\subsection{Charge transfer multiplet simulation}\label{sec:ctm}
In this section, we expand our investigation to explore how the change of electronic structure derived from the DFT+DMFT calculations is reflected in the shape of the Ni/Co 2$p$ photoemission spectra by utilizing the CTM model and comparing the results with experimental measurements.

As mentioned earlier, the CTM model defines $\Delta$ as the energy difference between the lowest $d^n$ and $d^{n+1}L^{-1}$ states (Fig.~\ref{fig:schema}(b) and (c)).\cite{bocquet1992electronic}
In specific, we define for the $d^n$ configuration, 
\begin{equation}
	n_d \epsilon_d+10*\epsilon_L+n_d(n_d-1) U_{dd} / 2=0,
	\label{eq:gs}
\end{equation}
and for the $d^{n+1}L^{-1}$ configuration, i.e, with an electron transferred from ligand to TM $d$ orbitals,
\begin{equation}
	(n_d+1) \epsilon_d+\left(10-1\right) \epsilon_L+n_d(n_d+1) U_{dd} / 2=\Delta.
	\label{eq:delta}
\end{equation}
Accordingly, the energy difference between $d^{n+1}L^{-1}$ and $d^{n+2}L^{-2}$ in the initial state can be derived, being $2\Delta+U_{dd}$, in which $U_{dd}$ denotes the Coulomb interaction within TM $d$ orbitals. In the final state, when the 2$p$ core electron is excited, it imposes a core-hole potential (described by the Coulomb interaction between TM 2$p$ core and 3$d$ valence states $U_{pd}$ ) on the initial configurations. 

\begin{figure}[h!]
	\centering
	\includegraphics[width=1.0\textwidth]{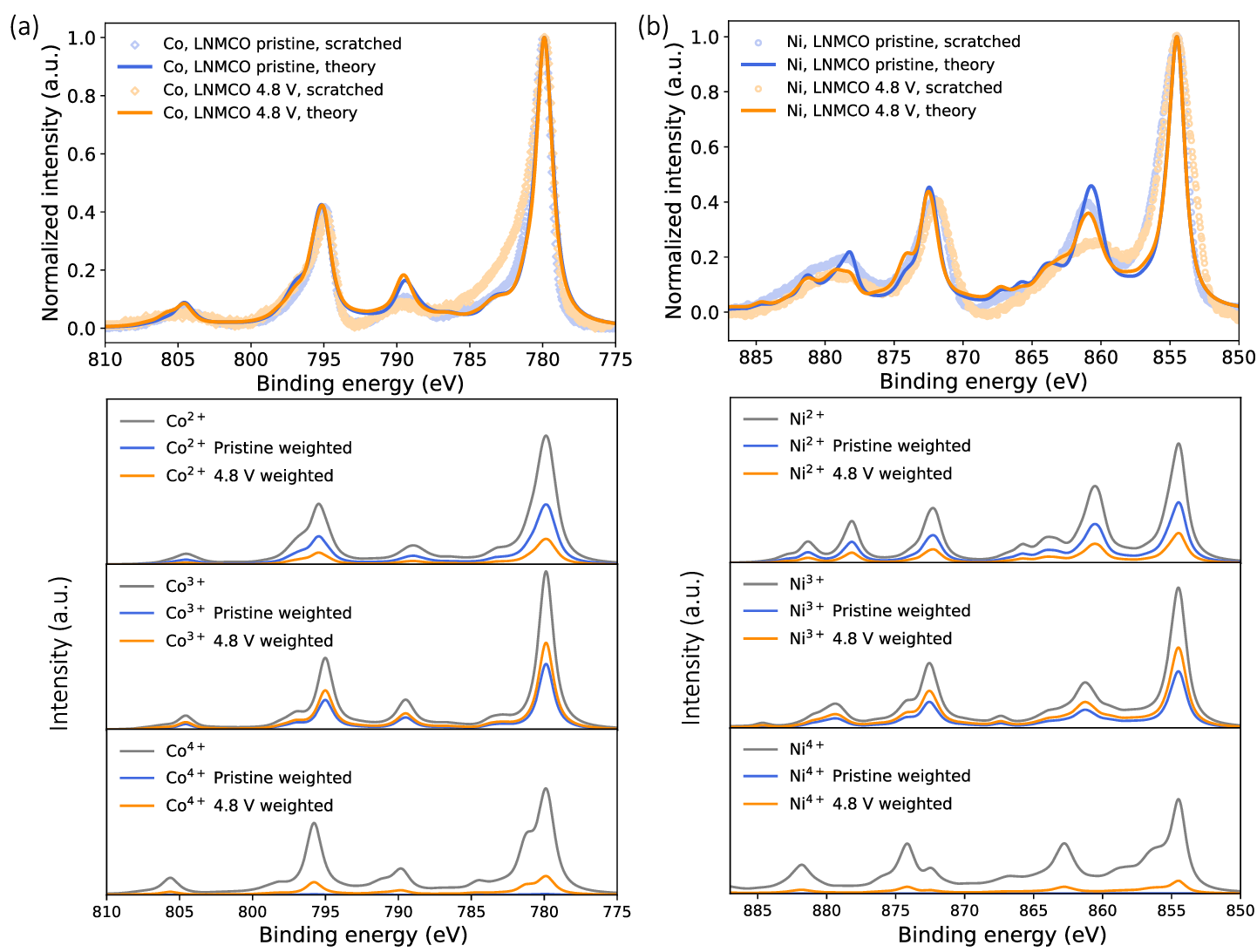}
	\caption{Simulated 2$p$ XPS of (a) Co and (b) Ni in LNMCO using Quanty based on the charge transfer multiplet (CTM) model. The final photoemission spectra for pristine and delithiated LNMCO are obtained by summing up the contributions from Co$^{2+}$/Ni$^{2+}$, Co$^{3+}$/Ni$^{3+}$, and Co$^{4+}$/Ni$^{4+}$ weighted by their corresponding probabilities as deduced by DMFT (see Fig.~\ref{fig:nmc_occu}). The experimental measurements are also shown for comparison.}
	\label{fig:ctm}
\end{figure}

Green and Sawatzky have classified correlated compounds into four categories based on the relative values of $U$ and $\Delta$: Mott-Hubbard ($U < \Delta$), charge transfer ($U > \Delta$) insulators, mixed valence and negative charge transfer regimes for small values of $\Delta$.\cite{green2024negative} Especially for correlated compounds which contain cations with high formal oxidation states, $\Delta$ can become small and even negative. In this regard, it is necessary that for cationic elements with different formal oxidation states, different values of $\Delta$ should be adopted in the CTM simulations.
Even though $\Delta$ describes a part of the electronic structure for the correlated compounds, the connection with the occupation number obtained by DMFT and the reference to individual oxidation states allow the application to mixed-TM multivalent compounds.
Following this line, we simulated the 2$p$ photoemission spectra of Co$^{2+}$/Ni$^{2+}$, Co$^{3+}$/Ni$^{3+}$, and Co$^{4+}$/Ni$^{4+}$ using the parameters listed in Tab.~\ref{tab:ctm_in}, in which the values of $\Delta$ for Co$^{2+}$/Ni$^{2+}$, Co$^{3+}$/Ni$^{3+}$, and Co$^{4+}$/Ni$^{4+}$ are selected based on the tables of collections for oxides as given in the work of Green and Sawatzky.\cite{green2024negative} It should be noted that the parameters which describe the crystal field splitting ($10D_q$ and $10D_q^L$) of Ni are adapted for different formal oxidation states so that there exists no abrupt low-spin to high-spin transition with higher oxidation states.  

The simulated 2$p$ photoemission spectra of Co$^{2+}$/Ni$^{2+}$, Co$^{3+}$/Ni$^{3+}$, and Co$^{4+}$/Ni$^{4+}$ are shown in Fig.~\ref{fig:ctm}. Together with their corresponding probabilities in pristine and delithiated LNMCO (see Fig.~\ref{fig:nmc_occu}), we arrive at the theoretical photoemission spectra taking the weighted sum. 
We note that the electronic configurations of Co and Ni, as obtained from DMFT for L$^{\mathrm{Co}}{1/3}$NMCO and L$^{\mathrm{Ni}}{1/3}$NMCO, can be used to simulate the fully delithiated systems.
The lowering of the 2$p$ XPS satellite peak intensity can be obviously observed for Ni, which well aligns with the experimentally measured XPS. Whereas for Co, the change with Li deintercalation is rather marginal. Such difference can be well understood from the individual XPS corresponding to Co$^{2+}$/Ni$^{2+}$, Co$^{3+}$/Ni$^{3+}$, and Co$^{4+}$/Ni$^{4+}$, respectively. Distinguished from Ni, for Co the satellite peak at around 790 eV is much weaker with respect to the main peak. The joint effect brought by lower proportion of Co$^{2+}$ and higher proportions of Co$^{3+}$ and Co$^{4+}$ is manifested only very weakly in the spectra. 
For Ni, the reduced proportion of Ni$^{2+}$ demonstrates a drastic decrease of its satellite peak, which is also clearly visible after taking the weighted sum (see Fig.~\ref{fig:ctm}(b)). 
No obvious spectral features which can be related to a formal Ni$^{4+}$ oxidation state are evident, however features to be assigned to a reduction of the formal Ni$^{2+}$ oxidation state are suggested.
Moreover, experimentally it has been found that the main 2$p$ XPS peak of Co gets broadened by delithiation, which is evidenced by the broadened main peak in the XPS corresponding to Co$^{4+}$. We expect that with increasing proportion of formal Co$^{4+}$, such broadening becomes more visible. 
As for Ni the experimental spectra of Co do not indicate the signature typical for the formation of a formal 4+ oxidation state. In the current CTM simulation, the hybridization of TM 3$d$ and O 2$p$ states is considered only for the neighbouring ligand states. Such cluster model can be extended further by including multiple ligand orbitals, as implemented in the LDA (local density approximation) + DMFT first-principle multiplet model.\cite{hariki2017lda+} We relate the observed difference between theory and experiment to the neglected non-local charge transfer effect, which, if considered, would lead to stronger hybridization effect, accompanied with the potentially increased $d$ ionization potential with increased positive charges due to Li deintercalation.

\section{Conclusions}
In conclusion, monitoring the evolution of the electronic structure in cathodes is crucial, as it directly influences the redox processes and overall performance of the battery material.
This study highlights the power of an integrated approach, combining experimental XPS measurements, DFT+DMFT calculations, and CTM simulations, to provide a comprehensive understanding of the evolution of the electronic structure of the cathodes during the battery's operation.
Based on the calculated hybridization functions using DFT+DMFT approach, the role of Mn in LiNi$_{1/3}$Mn$_{1/3}$Co$_{1/3}$O$_2$ as a system stabilizer is confirmed, as the 3$d$ states of Mn$^{4+}$ are not active in the hybridization. 
The 2$p$ XPS satellite peak intensity decreases with Li deintercalation, a phenomenon that can be attributed to the change of configurational distribution of TM 3$d$ states. In specific, the extraction of Li ions results in a reduced occupation number of the Co/Ni 3$d$ shell, consequently yielding a higher formal oxidation state. 
Additionally, stronger hybridization of TM 3$d$ states with O 2$p$ states coupled to hole transfer within the near-$E_F$ region is expected with deintercalation. 
As a consequence, a rigid band behaviour of hole transfer cannot be expected and O 2$p$ states that hybridize with TM 3$d$ states are more strongly involved in charge compensation (oxide redox).
However, as for the CTM simulations, the hybridization effect beyond single ligand charge transfer is not considered in the current work, the highly charged (strongly deintercalated) cathodes do not show the exact spectral structure to be expected for the higher formal oxidation states. 
We take this fact as additional evidence of TM 3$d$/O 2$p$ hybridization and the involvement of oxide redox processes.
From the detailed analysis of theoretical and experimental spectra, it is established that the satellite peak intensity in the TM 2$p$ core-level photoemission spectra serves as an effective indicator for the potential transition of TM  oxidation state, and thus providing evidence for the redox chemistry in cathode battery materials.


\section{Methodology}\label{sec:method}
\subsection{Experiment}
\subsubsection{Sample preparation}
LiCoO$_2$ (99.5\% pure, particle size: 5±2.5 $\mu$m, Alfa Aesar), LiNiO$_2$ (Sigma-Aldrich, 99.9\% purity),  LiNi$_{1/3}$Mn$_{1/3}$Co$_{1/3}$O$_2$ (LNMCO), (MSE Supplies, particle size: 7.5±2.5 $\mu$m D50, 99.9\% purity) powders were pressed into an aluminium mesh (Sigma-Aldrich, $\diameter=7 \mathrm{~mm}$) inside an argon-filled glovebox (MBraun, H$_2$O, and O$_2$ $<$ 0.1 ppm). The excess powder was removed from the rear of the mesh.
For the cathode discharged to 4.8 V, a fresh sample was put into a Swagelok cell to ensure good contact with the stainless-steel current collectors. A few drops of LP30 electrolyte (Sigma-Aldrich, 1.0 M LiPF$_6$ in EC/EMC=50/50, battery grade) and two separators (Celgard 2500) were introduced. Metallic lithium foil was used as an anode, which was scraped off to ensure a fresh surface. As shown in \inlinecite{mellin2024vacuo}, this set-up allows to obtain an electrochemical response as known from literature.
Before XPS characterization, the cell was disassembled, and the cathode samples were rinsed inside the glovebox with DMC (Sigma-Aldrich, $\geq$99\% purity). Rinsing was performed three times by spraying DMC onto the pressed powders using a syringe with a fine needle. The samples were transported to the XPS chamber without air exposure using a homemade transfer chamber and were stored in the UHV system overnight to allow the electrolyte to evaporate.

\subsubsection{XPS methodology}

All experiments were performed within a vacuum-cluster tool (Thermofisher ESCAlab 250, part of DAISY-SOL) with a pressure in the analysis chamber lower than 5$\times$10$^{-10}$ mbar. For excitation, monochromatic Al K$\alpha$ radiation (1486.6 eV) was used with a spot size of 650 $\mu$m. Detail spectra were measured by applying a pass energy of 25.0 eV and 0.1 eV/step. No charge neutralization was used. Binding energy calibration was performed regularly by setting the Au 4$f_{7/2}$ emission of Ar$^+$-cleaned Au foil to 84.00 eV. 

Backgrounds of the acquired spectra were subtracted using the Shirley method in the CasaXPS software (version 2.3.25) which is needed for the comparison to the simulated spectra. For Co 2p a background in the range of 775.8-792.8 eV and 792.8-809.9 eV and for Ni between 845.7-888.0 eV was used. It is important to note that the selection of background type and the energy can have an influence of the overall shape of the spectra. The spectra of the sample charged to 4.8 V was calibrated to the main peak of the pristine sample for Co and Ni due to the shifts to lower binding energy caused by the increase of work function.
The scratching process was performed by a movable scalpel installed inside a vacuum chamber with a base pressure lower than 5$\times$10$^{-7}$ mbar. By this \textit{in vacuo} scratching preparation, spectra with high signal to noise ratio was obtained to get bulk information of the samples made from the powered material.

\subsubsection{Electrochemical formation and charging}
The formation (to 4.2 V vs Li$^+$/Li) and the constant current (CC) plus constant voltage (CV) charging processes (to 4.8 V vs Li$^+$/Li) (CC: ±50 $\mu$A, CV-limit: ±5 $\mu$A) were performed with a Biologic VMP2 potentiostat in a temperature-controlled cabinet.\cite{mellin2024voltage} 

\subsection{DFT + DMFT method}
The stationary and charge self-consistent implementation of DFT + DMFT~\cite{haule2015free} is used to handle the many-body interaction in cathodes Li$_x$CoO$_2$, Li$_x$CoO$_2$, and Li$_x$Ni$_{1/3}$Mn$_{1/3}$Co$_{1/3}$O$_2$ ($x=1, 1/3$), where we treat Co, Ni and Mn 3$d$ orbitals as correlated sites. For the DFT part, the Kohn-Sham orbitals were solved using the WIEN2k package implementing a full-potential linear augmented plane-wave formalism.\cite{blaha2001wien2k} In this work the local density approximation (LDA) was employed as exchange-correlation functional. The cutoff R$_{\mathrm{MT}}$K$_{\mathrm{MAX}}$ was set to 7.0. The uniform k-mesh of $17 \times 17 \times 3$ was adopted for the Brillouin zone integration. The state-of-the-art continuous time quantum Monte Carlo (CTQMC) was used as the impurity solver.~\cite{haule2007quantum} 
We applied $U=8.0$ eV and $J=0.8$ eV for Co and Ni 3$d$ orbitals while for Mn 3$d$ orbitals, $U=9.0$ eV and $J=0.8$ eV were used. 
The selection of $U$ and $J$ for Co was first benchmarked by the calculated DOS of LiCoO$_2$~\cite{isaacs2020compositional}. For explicitness, the same $U$ and $J$ were used for Ni while for Mn the value of $U$ was slightly increase to 9 eV according to previous work on Mn oxides.\cite{mandal2019influence,mandal2019systematic} Furthermore, we have tested that the values of $U$ and $J$ in DMFT calculations will not change the trend observed during Li deintercalation.
The DMFT calculations were carried out at room temperature (equivalently $\beta = 38.7$) at paramagnetic state. In addition, the impurity problem was solved in local coordinates to ensure the maximal diagonalization of the hybridization function. The convergence criteria for charge and energy were $ 1 \times 10^{-5} e$ and $5 \times 10^{-6}$ Ry, respectively, which were tested to be sufficient for this work. The number of Monte Carlo steps for each impurity site at each iteration is $9.0 \times 10^7$.  The DOS was evaluated by implementing analytical continuations on the Matsubara self-energy functions $\Sigma(i\omega)$.

\subsection{Charge transfer multiplet model}
The QUANTY code~\cite{ackermann2024quanty} was used to solve the CTM model with its corresponding Hamiltonian defined as
\begin{equation}
H_{\mathrm{CTM}}=H_U^{d d}+H_U^{p d}+H_{l \cdot \mathrm{s}}^d+H_{l \cdot \mathrm{s}}^p+H_{on}^p+H_{on}^d+H_{on}^L+H_{\mathrm{hyb}}^{d L},
\label{eq:ctm}
\end{equation}
in which $H_U^{d d}$ denotes the on-site Coulomb repulsion between the $3 d$ electrons of the transition metal, $H_U^{p d}$ the on-site Coulomb interaction between the $2 p$ and $3 d$ electrons, $H_{l \cdot s}^d$ and $H_{l \cdot s}^p$ being the spin-orbital coupling corresponding to $3 d$ and $2 p$ orbitals, respectively. $H_{on}^p$, $H_{on}^d$ and $H_{on}^L$ denote the on-site energies of $2p$, $3d$ and the ligand orbitals, respectively. $H_{\mathrm{hyb}}^{d L}$ is then the hybridization matrix between transition metal $3d$ and the ligand orbitals. The input parameters used for describing the physical terms in Eq.~\ref{eq:ctm} are listed in Tab~\ref{tab:ctm_in}. The spectroscopy can then be calculated in accordance with Fermi's golden rule using the equation below:
\begin{equation}
G(\omega) = \langle\psi| T^{\dagger} \frac{1}{\left(\omega+\mathrm{i} \Gamma / 2+E_0-H_{CTM}^{f}\right)} T|\psi\rangle,
\label{eq:spectra}
\end{equation}
with $E_0=\langle\psi|H_{CTM}^{f}|\psi\rangle$.
Here, $|\psi\rangle$ denotes the ground state corresponding to the Hamiltonian defining the initial state, $T$ is the transition operator, $H_{CTM}^{f}$ is the Hamiltonian corresponding to the final state and $\Gamma$ indicates the broadening of the spectra. In this work, for simplicity the Gaussian and Lorentzian broadening were fixed to a FWHM = 0.4 and 1.0 eV over the whole energy range, respectively. 

\begin{table}[h!]
\centering
\caption{Parameters used in CTM model as derived for TM oxide.}
\label{tab:ctm_in}
\begin{tabular}{lcccccccc}
\hline \hline 
& $U_{dd}$ & $U_{pd}$ & $\Delta$ & $F_{dd}^2$ & $F_{dd}^4$ & $F_{pd}^2$ & $G_{pd}^1$ &  $G_{pd}^3$ \\
Co$^{2+}$ &  4.5 & 6.0  & 7.0 & 8.86 & 5.54 & 5.98 & 4.57 & 2.60 \\ 
Co$^{3+}$ &  4.5 & 6.0  & 4.0 & 8.86 & 5.54 & 5.98 & 4.57 & 2.60 \\ 
Co$^{4+}$ &  4.5 & 6.0  & -1.0 & 8.86 & 5.54 & 5.98 & 4.57 & 2.60 \\ 
Ni$^{2+}$ & 7.3 & 8.5 & 4.0 & 11.14 & 6.87 & 6.67 & 4.92 & 2.80 \\
Ni$^{3+}$ & 7.3 & 8.5 & 0.0 & 11.95 & 7.46 & 7.10 & 5.38 & 3.06 \\
Ni$^{4+}$ & 7.3 & 8.5 & -3.0 & 12.82 & 8.05 & 7.64 & 5.86 & 3.34 \\
\hline
& $10D_q$ & $10D_q^L$ & $V_{eg}$ & $V_{t2g}$ & $\xi_{3d}$ & $\xi_{2p}$ && \\
Co$^{2+}$ & 1.00 & 2.00 & 3.30 & 1.80 & 0.07 & 9.75 & & \\
Co$^{3+}$ & 1.00 & 2.00 & 3.30 & 1.80 & 0.07 & 9.75 & & \\
Co$^{3+}$ & 1.00 & 2.00 & 3.30 & 1.80 & 0.07 & 9.75 & & \\
Ni$^{2+}$  & 0.56 & 1.44 & 2.06 & 1.21 & 0.08 & 11.51 & & \\
Ni$^{3+}$  & 0.84 & 2.16 & 2.06 & 1.21 & 0.08 & 11.51 & & \\
Ni$^{4+}$  & 0.78 & 2.02 & 2.06 & 1.21 & 0.08 & 11.51 & & \\
\hline
\end{tabular}
\end{table}

\begin{acknowledgement}
The authors gratefully acknowledge the computing time provided to them at the NHR Center NHR4CES at RWTH Aachen University (project number p0020465 and p0024007) and at TU Darmstadt (project number p0020454). This is funded by the German Federal Ministry of Research, Technology, and Space (BMFTR), and the state governments participating on the basis of the resolutions of the Gemeinsame Wissenschaftskonferenz (GWK) for national high performance computing at universities. 
This work was also supported by the German Science Foundation (DFG, project number 416542991). In addition, we acknowledge the helpful discussions with Dr. Gennady Cherkashinin at TU Darmstadt.
\end{acknowledgement}

\bibliography{achemso-demo}

\providecommand{\latin}[1]{#1}
\makeatletter
\providecommand{\doi}
  {\begingroup\let\do\@makeother\dospecials
  \catcode`\{=1 \catcode`\}=2 \doi@aux}
\providecommand{\doi@aux}[1]{\endgroup\texttt{#1}}
\makeatother
\providecommand*\mcitethebibliography{\thebibliography}
\csname @ifundefined\endcsname{endmcitethebibliography}
  {\let\endmcitethebibliography\endthebibliography}{}
\begin{mcitethebibliography}{48}
\providecommand*\natexlab[1]{#1}
\providecommand*\mciteSetBstSublistMode[1]{}
\providecommand*\mciteSetBstMaxWidthForm[2]{}
\providecommand*\mciteBstWouldAddEndPuncttrue
  {\def\EndOfBibitem{\unskip.}}
\providecommand*\mciteBstWouldAddEndPunctfalse
  {\let\EndOfBibitem\relax}
\providecommand*\mciteSetBstMidEndSepPunct[3]{}
\providecommand*\mciteSetBstSublistLabelBeginEnd[3]{}
\providecommand*\EndOfBibitem{}
\mciteSetBstSublistMode{f}
\mciteSetBstMaxWidthForm{subitem}{(\alph{mcitesubitemcount})}
\mciteSetBstSublistLabelBeginEnd
  {\mcitemaxwidthsubitemform\space}
  {\relax}
  {\relax}

\bibitem[Reimers and Dahn(1992)Reimers, and Dahn]{reimers1992electrochemical}
Reimers,~J.~N.; Dahn,~J. Electrochemical and in situ {X}-ray diffraction
  studies of lithium intercalation in {L}i$_x${C}o{O}$_2$. \emph{Journal of the
  Electrochemical Society} \textbf{1992}, \emph{139}, 2091\relax
\mciteBstWouldAddEndPuncttrue
\mciteSetBstMidEndSepPunct{\mcitedefaultmidpunct}
{\mcitedefaultendpunct}{\mcitedefaultseppunct}\relax
\EndOfBibitem
\bibitem[Wang \latin{et~al.}(2004)Wang, Sun, Chen, and
  Huang]{wang2004electrochemical}
Wang,~Z.; Sun,~Y.; Chen,~L.; Huang,~X. Electrochemical characterization of
  positive electrode material {L}i{N}i$_{1/3}${C}o$_{1/3}${M}n$_{1/3}$O$_2$ and
  compatibility with electrolyte for lithium-ion batteries. \emph{Journal of
  the Electrochemical Society} \textbf{2004}, \emph{151}, A914\relax
\mciteBstWouldAddEndPuncttrue
\mciteSetBstMidEndSepPunct{\mcitedefaultmidpunct}
{\mcitedefaultendpunct}{\mcitedefaultseppunct}\relax
\EndOfBibitem
\bibitem[Gauthier \latin{et~al.}(2018)Gauthier, Karayaylali, Giordano, Feng,
  Lux, Maglia, Lamp, and Shao-Horn]{gauthier2018probing}
Gauthier,~M.; Karayaylali,~P.; Giordano,~L.; Feng,~S.; Lux,~S.~F.; Maglia,~F.;
  Lamp,~P.; Shao-Horn,~Y. Probing surface chemistry changes using
  {L}i{C}oO$_2$-only electrodes in {L}i-ion batteries. \emph{Journal of The
  Electrochemical Society} \textbf{2018}, \emph{165}, A1377\relax
\mciteBstWouldAddEndPuncttrue
\mciteSetBstMidEndSepPunct{\mcitedefaultmidpunct}
{\mcitedefaultendpunct}{\mcitedefaultseppunct}\relax
\EndOfBibitem
\bibitem[Schulz \latin{et~al.}(2018)Schulz, Hausbrand, Dimesso, and
  Jaegermann]{schulz2018xps}
Schulz,~N.; Hausbrand,~R.; Dimesso,~L.; Jaegermann,~W. {XPS}-surface analysis
  of {SEI} layers on {L}i-ion cathodes: {P}art {I}. {I}nvestigation of initial
  surface chemistry. \emph{Journal of The Electrochemical Society}
  \textbf{2018}, \emph{165}, A819\relax
\mciteBstWouldAddEndPuncttrue
\mciteSetBstMidEndSepPunct{\mcitedefaultmidpunct}
{\mcitedefaultendpunct}{\mcitedefaultseppunct}\relax
\EndOfBibitem
\bibitem[Hausbrand(2020)]{hausbrand2020electronic}
Hausbrand,~R. Electronic energy levels at {L}i-ion cathode--liquid electrolyte
  interfaces: {C}oncepts, experimental insights, and perspectives. \emph{The
  Journal of Chemical Physics} \textbf{2020}, \emph{152}\relax
\mciteBstWouldAddEndPuncttrue
\mciteSetBstMidEndSepPunct{\mcitedefaultmidpunct}
{\mcitedefaultendpunct}{\mcitedefaultseppunct}\relax
\EndOfBibitem
\bibitem[Hausbrand \latin{et~al.}(2017)Hausbrand, Cherkashinin, Fingerle, and
  Jaegermann]{hausbrand2017surface}
Hausbrand,~R.; Cherkashinin,~G.; Fingerle,~M.; Jaegermann,~W. Surface and bulk
  properties of {L}i-ion electrodes--{A} surface science approach.
  \emph{Journal of Electron Spectroscopy and Related Phenomena} \textbf{2017},
  \emph{221}, 65--78\relax
\mciteBstWouldAddEndPuncttrue
\mciteSetBstMidEndSepPunct{\mcitedefaultmidpunct}
{\mcitedefaultendpunct}{\mcitedefaultseppunct}\relax
\EndOfBibitem
\bibitem[Ensling \latin{et~al.}(2014)Ensling, Cherkashinin, Schmid,
  Bhuvaneswari, Thissen, and Jaegermann]{ensling2014nonrigid}
Ensling,~D.; Cherkashinin,~G.; Schmid,~S.; Bhuvaneswari,~S.; Thissen,~A.;
  Jaegermann,~W. Nonrigid band behavior of the electronic structure of
  $\mathrm{LiCoO}_2$ thin film during electrochemical {L}i deintercalation.
  \emph{Chemistry of Materials} \textbf{2014}, \emph{26}, 3948--3956\relax
\mciteBstWouldAddEndPuncttrue
\mciteSetBstMidEndSepPunct{\mcitedefaultmidpunct}
{\mcitedefaultendpunct}{\mcitedefaultseppunct}\relax
\EndOfBibitem
\bibitem[Naylor \latin{et~al.}(2019)Naylor, Makkos, Maibach, Guerrini,
  Sobkowiak, Bj{\"o}rklund, Lozano, Menon, Younesi, Roberts, \latin{et~al.}
  others]{naylor2019depth}
Naylor,~A.~J.; Makkos,~E.; Maibach,~J.; Guerrini,~N.; Sobkowiak,~A.;
  Bj{\"o}rklund,~E.; Lozano,~J.~G.; Menon,~A.~S.; Younesi,~R.; Roberts,~M.~R.,
  \latin{et~al.}  Depth-dependent oxygen redox activity in lithium-rich layered
  oxide cathodes. \emph{Journal of Materials Chemistry A} \textbf{2019},
  \emph{7}, 25355--25368\relax
\mciteBstWouldAddEndPuncttrue
\mciteSetBstMidEndSepPunct{\mcitedefaultmidpunct}
{\mcitedefaultendpunct}{\mcitedefaultseppunct}\relax
\EndOfBibitem
\bibitem[Flores \latin{et~al.}(2021)Flores, Mozhzhukhina, Aschauer, and
  Berg]{flores2021operando}
Flores,~E.; Mozhzhukhina,~N.; Aschauer,~U.; Berg,~E.~J. Operando monitoring the
  insulator--metal transition of $\mathrm{LiCoO}_2$. \emph{ACS Applied
  Materials \& Interfaces} \textbf{2021}, \emph{13}, 22540--22548\relax
\mciteBstWouldAddEndPuncttrue
\mciteSetBstMidEndSepPunct{\mcitedefaultmidpunct}
{\mcitedefaultendpunct}{\mcitedefaultseppunct}\relax
\EndOfBibitem
\bibitem[Uchimoto \latin{et~al.}(2001)Uchimoto, Sawada, and
  Yao]{uchimoto2001changes}
Uchimoto,~Y.; Sawada,~H.; Yao,~T. Changes in electronic structure by {L}i ion
  deintercalation in $\mathrm{LiCoO}_2$ from cobalt {L}-edge and oxygen
  {K}-edge {XANES}. \emph{Journal of Synchrotron Radiation} \textbf{2001},
  \emph{8}, 872--873\relax
\mciteBstWouldAddEndPuncttrue
\mciteSetBstMidEndSepPunct{\mcitedefaultmidpunct}
{\mcitedefaultendpunct}{\mcitedefaultseppunct}\relax
\EndOfBibitem
\bibitem[Jung \latin{et~al.}(2017)Jung, Metzger, Maglia, Stinner, and
  Gasteiger]{jung2017oxygen}
Jung,~R.; Metzger,~M.; Maglia,~F.; Stinner,~C.; Gasteiger,~H.~A. Oxygen release
  and its effect on the cycling stability of
  {L}i{N}i$_x${M}n$_y${C}o$_z${O}$_2$ ({NMC}) cathode materials for {L}i-ion
  batteries. \emph{Journal of The Electrochemical Society} \textbf{2017},
  \emph{164}, A1361\relax
\mciteBstWouldAddEndPuncttrue
\mciteSetBstMidEndSepPunct{\mcitedefaultmidpunct}
{\mcitedefaultendpunct}{\mcitedefaultseppunct}\relax
\EndOfBibitem
\bibitem[Li \latin{et~al.}(2022)Li, Ning, Wong, An, Tang, Zhou, Schuck, Chen,
  Zhang, and Liu]{li2022improving}
Li,~Q.; Ning,~D.; Wong,~D.; An,~K.; Tang,~Y.; Zhou,~D.; Schuck,~G.; Chen,~Z.;
  Zhang,~N.; Liu,~X. Improving the oxygen redox reversibility of {L}i-rich
  battery cathode materials via {C}oulombic repulsive interactions strategy.
  \emph{Nature Communications} \textbf{2022}, \emph{13}, 1123\relax
\mciteBstWouldAddEndPuncttrue
\mciteSetBstMidEndSepPunct{\mcitedefaultmidpunct}
{\mcitedefaultendpunct}{\mcitedefaultseppunct}\relax
\EndOfBibitem
\bibitem[Li \latin{et~al.}(2021)Li, Liu, Wang, Feng, Ren, Wu, Xu, Lu, Hou,
  Zhang, \latin{et~al.} others]{li2021thermal}
Li,~Y.; Liu,~X.; Wang,~L.; Feng,~X.; Ren,~D.; Wu,~Y.; Xu,~G.; Lu,~L.; Hou,~J.;
  Zhang,~W., \latin{et~al.}  Thermal runaway mechanism of lithium-ion battery
  with {L}i{N}i$_{0.8}${M}n$_{0.1}${C}o$_{0.1}${O}$_2$ cathode materials.
  \emph{Nano Energy} \textbf{2021}, \emph{85}, 105878\relax
\mciteBstWouldAddEndPuncttrue
\mciteSetBstMidEndSepPunct{\mcitedefaultmidpunct}
{\mcitedefaultendpunct}{\mcitedefaultseppunct}\relax
\EndOfBibitem
\bibitem[Assat and Tarascon(2018)Assat, and Tarascon]{assat2018fundamental}
Assat,~G.; Tarascon,~J.-M. Fundamental understanding and practical challenges
  of anionic redox activity in {L}i-ion batteries. \emph{Nature Energy}
  \textbf{2018}, \emph{3}, 373--386\relax
\mciteBstWouldAddEndPuncttrue
\mciteSetBstMidEndSepPunct{\mcitedefaultmidpunct}
{\mcitedefaultendpunct}{\mcitedefaultseppunct}\relax
\EndOfBibitem
\bibitem[Ben~Yahia \latin{et~al.}(2019)Ben~Yahia, Vergnet, Sauban{\`e}re, and
  Doublet]{ben2019unified}
Ben~Yahia,~M.; Vergnet,~J.; Sauban{\`e}re,~M.; Doublet,~M.-L. Unified picture
  of anionic redox in {L}i/{N}a-ion batteries. \emph{Nature Materials}
  \textbf{2019}, \emph{18}, 496--502\relax
\mciteBstWouldAddEndPuncttrue
\mciteSetBstMidEndSepPunct{\mcitedefaultmidpunct}
{\mcitedefaultendpunct}{\mcitedefaultseppunct}\relax
\EndOfBibitem
\bibitem[House \latin{et~al.}(2021)House, Marie, P{\'e}rez-Osorio, Rees,
  Boivin, and Bruce]{house2021role}
House,~R.~A.; Marie,~J.-J.; P{\'e}rez-Osorio,~M.~A.; Rees,~G.~J.; Boivin,~E.;
  Bruce,~P.~G. The role of $\mathrm{O}_2$ in {O}-redox cathodes for {L}i-ion
  batteries. \emph{Nature Energy} \textbf{2021}, \emph{6}, 781--789\relax
\mciteBstWouldAddEndPuncttrue
\mciteSetBstMidEndSepPunct{\mcitedefaultmidpunct}
{\mcitedefaultendpunct}{\mcitedefaultseppunct}\relax
\EndOfBibitem
\bibitem[Menon \latin{et~al.}(2023)Menon, Johnston, Booth, Zhang, Kress,
  Murdock, Paez~Fajardo, Anthonisamy, Tapia-Ruiz, Agrestini, \latin{et~al.}
  others]{menon2023oxygen}
Menon,~A.~S.; Johnston,~B.~J.; Booth,~S.~G.; Zhang,~L.; Kress,~K.; Murdock,~B.;
  Paez~Fajardo,~G.; Anthonisamy,~N.~N.; Tapia-Ruiz,~N.; Agrestini,~S.,
  \latin{et~al.}  Oxygen-redox activity in non-lithium-excess tungsten-doped
  $\mathrm{LiNiO}_2$ cathode. \emph{PRX Energy} \textbf{2023}, \emph{2},
  013005\relax
\mciteBstWouldAddEndPuncttrue
\mciteSetBstMidEndSepPunct{\mcitedefaultmidpunct}
{\mcitedefaultendpunct}{\mcitedefaultseppunct}\relax
\EndOfBibitem
\bibitem[Van~Elp \latin{et~al.}(1991)Van~Elp, Wieland, Eskes, Kuiper, Sawatzky,
  De~Groot, and Turner]{van1991electronic}
Van~Elp,~J.; Wieland,~J.; Eskes,~H.; Kuiper,~P.; Sawatzky,~G.; De~Groot,~F.;
  Turner,~T. Electronic structure of $\mathrm{CoO}$, {L}i-doped $\mathrm{CoO}$,
  and $\mathrm{LiCoO}_2$. \emph{Physical Review B} \textbf{1991}, \emph{44},
  6090\relax
\mciteBstWouldAddEndPuncttrue
\mciteSetBstMidEndSepPunct{\mcitedefaultmidpunct}
{\mcitedefaultendpunct}{\mcitedefaultseppunct}\relax
\EndOfBibitem
\bibitem[Liang \latin{et~al.}(2025)Liang, Baubaid, Radtke, Mellin, Maheu,
  Maiti, Sclar, P{\'\i}{\v{s}}, Nappini, Magnano, \latin{et~al.}
  others]{liang2025novel}
Liang,~Z.; Baubaid,~A.; Radtke,~M.; Mellin,~M.; Maheu,~C.; Maiti,~S.;
  Sclar,~H.; P{\'\i}{\v{s}},~I.; Nappini,~S.; Magnano,~E., \latin{et~al.}
  Novel Insights into Enhanced Stability of Li-Rich Layered and High-Voltage
  Olivine Phosphate Cathodes for Advanced Batteries through Surface
  Modification and Electron Structure Design. \emph{Advanced Science}
  \textbf{2025}, \emph{12}, 2413054\relax
\mciteBstWouldAddEndPuncttrue
\mciteSetBstMidEndSepPunct{\mcitedefaultmidpunct}
{\mcitedefaultendpunct}{\mcitedefaultseppunct}\relax
\EndOfBibitem
\bibitem[Kuiper \latin{et~al.}(1989)Kuiper, Kruizinga, Ghijsen, Sawatzky, and
  Verweij]{kuiper1989character}
Kuiper,~P.; Kruizinga,~G.; Ghijsen,~J.; Sawatzky,~G.~A.; Verweij,~H. Character
  of holes in $\mathrm{Li_xNi_{1-x}O}$ and their magnetic behavior.
  \emph{Physical Review Letters} \textbf{1989}, \emph{62}, 221\relax
\mciteBstWouldAddEndPuncttrue
\mciteSetBstMidEndSepPunct{\mcitedefaultmidpunct}
{\mcitedefaultendpunct}{\mcitedefaultseppunct}\relax
\EndOfBibitem
\bibitem[Genreith-Schriever \latin{et~al.}(2023)Genreith-Schriever, Banerjee,
  Menon, Bassey, Piper, Grey, and Morris]{genreith2023oxygen}
Genreith-Schriever,~A.~R.; Banerjee,~H.; Menon,~A.~S.; Bassey,~E.~N.;
  Piper,~L.~F.; Grey,~C.~P.; Morris,~A.~J. Oxygen hole formation controls
  stability in $\mathrm{LiNiO}_2$ cathodes. \emph{Joule} \textbf{2023},
  \emph{7}, 1623--1640\relax
\mciteBstWouldAddEndPuncttrue
\mciteSetBstMidEndSepPunct{\mcitedefaultmidpunct}
{\mcitedefaultendpunct}{\mcitedefaultseppunct}\relax
\EndOfBibitem
\bibitem[De~Groot(2005)]{de2005multiplet}
De~Groot,~F. Multiplet effects in {X}-ray spectroscopy. \emph{Coordination
  Chemistry Reviews} \textbf{2005}, \emph{249}, 31--63\relax
\mciteBstWouldAddEndPuncttrue
\mciteSetBstMidEndSepPunct{\mcitedefaultmidpunct}
{\mcitedefaultendpunct}{\mcitedefaultseppunct}\relax
\EndOfBibitem
\bibitem[Dah{\'e}ron \latin{et~al.}(2008)Dah{\'e}ron, Dedryv{\`e}re, Martinez,
  M{\'e}n{\'e}trier, Denage, Delmas, and Gonbeau]{daheron2008electron}
Dah{\'e}ron,~L.; Dedryv{\`e}re,~R.; Martinez,~H.; M{\'e}n{\'e}trier,~M.;
  Denage,~C.; Delmas,~C.; Gonbeau,~D. Electron transfer mechanisms upon lithium
  deintercalation from {L}i{C}o{O}$_2$ to {C}o{O}$_2$ investigated by {XPS}.
  \emph{Chemistry of Materials} \textbf{2008}, \emph{20}, 583--590\relax
\mciteBstWouldAddEndPuncttrue
\mciteSetBstMidEndSepPunct{\mcitedefaultmidpunct}
{\mcitedefaultendpunct}{\mcitedefaultseppunct}\relax
\EndOfBibitem
\bibitem[Zhang \latin{et~al.}(2023)Zhang, Xie, Long, G{\"u}nzing, Wende,
  Ollefs, and Zhang]{zhang2023autonomous}
Zhang,~Y.; Xie,~R.; Long,~T.; G{\"u}nzing,~D.; Wende,~H.; Ollefs,~K.~J.;
  Zhang,~H. Autonomous atomic Hamiltonian construction and active sampling of
  {X}-ray absorption spectroscopy by adversarial {B}ayesian optimization.
  \emph{npj Computational Materials} \textbf{2023}, \emph{9}, 46\relax
\mciteBstWouldAddEndPuncttrue
\mciteSetBstMidEndSepPunct{\mcitedefaultmidpunct}
{\mcitedefaultendpunct}{\mcitedefaultseppunct}\relax
\EndOfBibitem
\bibitem[Isaacs and Marianetti(2020)Isaacs, and
  Marianetti]{isaacs2020compositional}
Isaacs,~E.~B.; Marianetti,~C.~A. Compositional phase stability of correlated
  electron materials within {DFT} + {DMFT}. \emph{Physical Review B}
  \textbf{2020}, \emph{102}, 045146\relax
\mciteBstWouldAddEndPuncttrue
\mciteSetBstMidEndSepPunct{\mcitedefaultmidpunct}
{\mcitedefaultendpunct}{\mcitedefaultseppunct}\relax
\EndOfBibitem
\bibitem[Fantin \latin{et~al.}(2023)Fantin, van Roekeghem, and
  Benayad]{fantin2023self}
Fantin,~R.; van Roekeghem,~A.; Benayad,~A. Self-Regulated Ligand-Metal Charge
  Transfer upon Lithium-Ion Deintercalation Process from $\mathrm{LiCoO}_2$ to
  $\mathrm{CoO}_2$. \emph{PRX Energy} \textbf{2023}, \emph{2}, 043010\relax
\mciteBstWouldAddEndPuncttrue
\mciteSetBstMidEndSepPunct{\mcitedefaultmidpunct}
{\mcitedefaultendpunct}{\mcitedefaultseppunct}\relax
\EndOfBibitem
\bibitem[Nishizawa and Yamamura(1998)Nishizawa, and
  Yamamura]{nishizawa1998irreversible}
Nishizawa,~M.; Yamamura,~S. Irreversible conductivity change of
  {L}i$_{1-x}${C}o{O}$_2$ on electrochemical lithium insertion/extraction,
  desirable for battery applications. \emph{Chemical Communications}
  \textbf{1998}, 1631--1632\relax
\mciteBstWouldAddEndPuncttrue
\mciteSetBstMidEndSepPunct{\mcitedefaultmidpunct}
{\mcitedefaultendpunct}{\mcitedefaultseppunct}\relax
\EndOfBibitem
\bibitem[Cherkashinin \latin{et~al.}(2015)Cherkashinin, Motzko, Schulz,
  Sp\''{a}th, and Jaegermann]{cherkashinin2015electron}
Cherkashinin,~G.; Motzko,~M.; Schulz,~N.; Sp\''{a}th,~T.; Jaegermann,~W.
  Electron spectroscopy study of $\mathrm{Li[Ni,Co,Mn]O}_2$/electrolyte
  interface: electronic structure, interface composition, and device
  implications. \emph{Chemistry of Materials} \textbf{2015}, \emph{27},
  2875--2887\relax
\mciteBstWouldAddEndPuncttrue
\mciteSetBstMidEndSepPunct{\mcitedefaultmidpunct}
{\mcitedefaultendpunct}{\mcitedefaultseppunct}\relax
\EndOfBibitem
\bibitem[Manthiram(2020)]{manthiram2020reflection}
Manthiram,~A. A reflection on lithium-ion battery cathode chemistry.
  \emph{Nature Communications} \textbf{2020}, \emph{11}, 1550\relax
\mciteBstWouldAddEndPuncttrue
\mciteSetBstMidEndSepPunct{\mcitedefaultmidpunct}
{\mcitedefaultendpunct}{\mcitedefaultseppunct}\relax
\EndOfBibitem
\bibitem[Bisogni \latin{et~al.}(2016)Bisogni, Catalano, Green, Gibert,
  Scherwitzl, Huang, Strocov, Zubko, Balandeh, Triscone, \latin{et~al.}
  others]{bisogni2016ground}
Bisogni,~V.; Catalano,~S.; Green,~R.~J.; Gibert,~M.; Scherwitzl,~R.; Huang,~Y.;
  Strocov,~V.~N.; Zubko,~P.; Balandeh,~S.; Triscone,~J.-M., \latin{et~al.}
  Ground-state oxygen holes and the metal--insulator transition in the negative
  charge-transfer rare-earth nickelates. \emph{Nature Communications}
  \textbf{2016}, \emph{7}, 13017\relax
\mciteBstWouldAddEndPuncttrue
\mciteSetBstMidEndSepPunct{\mcitedefaultmidpunct}
{\mcitedefaultendpunct}{\mcitedefaultseppunct}\relax
\EndOfBibitem
\bibitem[Pavarini \latin{et~al.}(2016)Pavarini, Koch, van~den Brink, and
  Sawatzky]{pavarini2016quantum}
Pavarini,~E.; Koch,~E.; van~den Brink,~J.; Sawatzky,~G. Quantum Materials:
  {E}xperiments and Theory. \emph{Modeling and Simulation} \textbf{2016},
  \emph{6}, 420\relax
\mciteBstWouldAddEndPuncttrue
\mciteSetBstMidEndSepPunct{\mcitedefaultmidpunct}
{\mcitedefaultendpunct}{\mcitedefaultseppunct}\relax
\EndOfBibitem
\bibitem[Mellin \latin{et~al.}(2024)Mellin, Cherkashinin, Mohseni, Phillips,
  Jaegermann, and Hofmann]{mellin2024voltage}
Mellin,~M.; Cherkashinin,~G.; Mohseni,~E.; Phillips,~R.; Jaegermann,~W.;
  Hofmann,~J.~P. Voltage-dependent charge compensation mechanism and cathode
  electrolyte interface stability of the lithium-ion battery cathode materials
  $\mathrm{LiCoO_2}$ and $\mathrm{LiNi_{1/3}Mn_{1/3}Co_{1/3}O_2}$ studied by
  photoelectron spectroscopy. \emph{Journal of Materials Chemistry A}
  \textbf{2024}, \emph{12}, 3644--3658\relax
\mciteBstWouldAddEndPuncttrue
\mciteSetBstMidEndSepPunct{\mcitedefaultmidpunct}
{\mcitedefaultendpunct}{\mcitedefaultseppunct}\relax
\EndOfBibitem
\bibitem[Zaanen \latin{et~al.}(1985)Zaanen, Sawatzky, and
  Allen]{zaanen1985band}
Zaanen,~J.; Sawatzky,~G.; Allen,~J. Band gaps and electronic structure of
  transition-metal compounds. \emph{Physical Review Letters} \textbf{1985},
  \emph{55}, 418\relax
\mciteBstWouldAddEndPuncttrue
\mciteSetBstMidEndSepPunct{\mcitedefaultmidpunct}
{\mcitedefaultendpunct}{\mcitedefaultseppunct}\relax
\EndOfBibitem
\bibitem[Van~Veenendaal and Sawatzky(1993)Van~Veenendaal, and
  Sawatzky]{van1993nonlocal}
Van~Veenendaal,~M.; Sawatzky,~G. Nonlocal screening effects in 2p x-ray
  photoemission spectroscopy core-level line shapes of transition metal
  compounds. \emph{Physical Review Letters} \textbf{1993}, \emph{70},
  2459\relax
\mciteBstWouldAddEndPuncttrue
\mciteSetBstMidEndSepPunct{\mcitedefaultmidpunct}
{\mcitedefaultendpunct}{\mcitedefaultseppunct}\relax
\EndOfBibitem
\bibitem[Altieri \latin{et~al.}(2000)Altieri, Tjeng, Tanaka, and
  Sawatzky]{altieri2000core}
Altieri,~S.; Tjeng,~L.; Tanaka,~A.; Sawatzky,~G. Core-level x-ray photoemission
  on {N}i{O} in the impurity limit. \emph{Physical Review B} \textbf{2000},
  \emph{61}, 13403\relax
\mciteBstWouldAddEndPuncttrue
\mciteSetBstMidEndSepPunct{\mcitedefaultmidpunct}
{\mcitedefaultendpunct}{\mcitedefaultseppunct}\relax
\EndOfBibitem
\bibitem[Green \latin{et~al.}(2016)Green, Haverkort, and
  Sawatzky]{green2016bond}
Green,~R.~J.; Haverkort,~M.~W.; Sawatzky,~G.~A. Bond disproportionation and
  dynamical charge fluctuations in the perovskite rare-earth nickelates.
  \emph{Physical Review B} \textbf{2016}, \emph{94}, 195127\relax
\mciteBstWouldAddEndPuncttrue
\mciteSetBstMidEndSepPunct{\mcitedefaultmidpunct}
{\mcitedefaultendpunct}{\mcitedefaultseppunct}\relax
\EndOfBibitem
\bibitem[Dah{\'e}ron \latin{et~al.}(2009)Dah{\'e}ron, Martinez, Dedryv{\`e}re,
  Baraille, M{\'e}n{\'e}trier, Denage, Delmas, and Gonbeau]{daheron2009surface}
Dah{\'e}ron,~L.; Martinez,~H.; Dedryv{\`e}re,~R.; Baraille,~I.;
  M{\'e}n{\'e}trier,~M.; Denage,~C.; Delmas,~C.; Gonbeau,~D. Surface properties
  of {L}i{C}o{O}$_2$ investigated by {XPS} analyses and theoretical
  calculations. \emph{The Journal of Physical Chemistry C} \textbf{2009},
  \emph{113}, 5843--5852\relax
\mciteBstWouldAddEndPuncttrue
\mciteSetBstMidEndSepPunct{\mcitedefaultmidpunct}
{\mcitedefaultendpunct}{\mcitedefaultseppunct}\relax
\EndOfBibitem
\bibitem[Bocquet \latin{et~al.}(1992)Bocquet, Mizokawa, Saitoh, Namatame, and
  Fujimori]{bocquet1992electronic}
Bocquet,~A.; Mizokawa,~T.; Saitoh,~T.; Namatame,~H.; Fujimori,~A. Electronic
  structure of 3d-transition-metal compounds by analysis of the 2p core-level
  photoemission spectra. \emph{Physical Review B} \textbf{1992}, \emph{46},
  3771\relax
\mciteBstWouldAddEndPuncttrue
\mciteSetBstMidEndSepPunct{\mcitedefaultmidpunct}
{\mcitedefaultendpunct}{\mcitedefaultseppunct}\relax
\EndOfBibitem
\bibitem[Green and Sawatzky(2024)Green, and Sawatzky]{green2024negative}
Green,~R.~J.; Sawatzky,~G.~A. Negative charge transfer energy in correlated
  compounds. \emph{Journal of the Physical Society of Japan} \textbf{2024},
  \emph{93}, 121007\relax
\mciteBstWouldAddEndPuncttrue
\mciteSetBstMidEndSepPunct{\mcitedefaultmidpunct}
{\mcitedefaultendpunct}{\mcitedefaultseppunct}\relax
\EndOfBibitem
\bibitem[Hariki \latin{et~al.}(2017)Hariki, Uozumi, and
  Kune{\v{s}}]{hariki2017lda+}
Hariki,~A.; Uozumi,~T.; Kune{\v{s}},~J. {LDA} + {DMFT} approach to core-level
  spectroscopy: {A}pplication to 3d transition metal compounds. \emph{Physical
  Review B} \textbf{2017}, \emph{96}, 045111\relax
\mciteBstWouldAddEndPuncttrue
\mciteSetBstMidEndSepPunct{\mcitedefaultmidpunct}
{\mcitedefaultendpunct}{\mcitedefaultseppunct}\relax
\EndOfBibitem
\bibitem[Mellin \latin{et~al.}(2024)Mellin, Cherkashinin, Jaegermann, and
  Hofmann]{mellin2024vacuo}
Mellin,~M.; Cherkashinin,~G.; Jaegermann,~W.; Hofmann,~J.~P. In Vacuo
  Scratching Yields Undisturbed Insight into the Bulk of Lithium-Ion Battery
  Positive Electrode Materials. \emph{ACS Energy Letters} \textbf{2024},
  \emph{9}, 4922--4928\relax
\mciteBstWouldAddEndPuncttrue
\mciteSetBstMidEndSepPunct{\mcitedefaultmidpunct}
{\mcitedefaultendpunct}{\mcitedefaultseppunct}\relax
\EndOfBibitem
\bibitem[Haule and Birol(2015)Haule, and Birol]{haule2015free}
Haule,~K.; Birol,~T. Free energy from stationary implementation of the {DFT} +
  {DMFT} functional. \emph{Physical Review Letters} \textbf{2015}, \emph{115},
  256402\relax
\mciteBstWouldAddEndPuncttrue
\mciteSetBstMidEndSepPunct{\mcitedefaultmidpunct}
{\mcitedefaultendpunct}{\mcitedefaultseppunct}\relax
\EndOfBibitem
\bibitem[Blaha \latin{et~al.}(2001)Blaha, Schwarz, Madsen, Kvasnicka, Luitz,
  \latin{et~al.} others]{blaha2001wien2k}
Blaha,~P.; Schwarz,~K.; Madsen,~G.~K.; Kvasnicka,~D.; Luitz,~J., \latin{et~al.}
   wien2k. \emph{An augmented plane wave + local orbitals program for
  calculating crystal properties} \textbf{2001}, \emph{60}\relax
\mciteBstWouldAddEndPuncttrue
\mciteSetBstMidEndSepPunct{\mcitedefaultmidpunct}
{\mcitedefaultendpunct}{\mcitedefaultseppunct}\relax
\EndOfBibitem
\bibitem[Haule(2007)]{haule2007quantum}
Haule,~K. Quantum Monte Carlo impurity solver for cluster dynamical mean-field
  theory and electronic structure calculations with adjustable cluster base.
  \emph{Physical Review B—Condensed Matter and Materials Physics}
  \textbf{2007}, \emph{75}, 155113\relax
\mciteBstWouldAddEndPuncttrue
\mciteSetBstMidEndSepPunct{\mcitedefaultmidpunct}
{\mcitedefaultendpunct}{\mcitedefaultseppunct}\relax
\EndOfBibitem
\bibitem[Mandal \latin{et~al.}(2019)Mandal, Haule, Rabe, and
  Vanderbilt]{mandal2019influence}
Mandal,~S.; Haule,~K.; Rabe,~K.~M.; Vanderbilt,~D. Influence of magnetic
  ordering on the spectral properties of binary transition metal oxides.
  \emph{Physical Review B} \textbf{2019}, \emph{100}, 245109\relax
\mciteBstWouldAddEndPuncttrue
\mciteSetBstMidEndSepPunct{\mcitedefaultmidpunct}
{\mcitedefaultendpunct}{\mcitedefaultseppunct}\relax
\EndOfBibitem
\bibitem[Mandal \latin{et~al.}(2019)Mandal, Haule, Rabe, and
  Vanderbilt]{mandal2019systematic}
Mandal,~S.; Haule,~K.; Rabe,~K.~M.; Vanderbilt,~D. Systematic beyond-{DFT}
  study of binary transition metal oxides. \emph{npj Computational Materials}
  \textbf{2019}, \emph{5}, 115\relax
\mciteBstWouldAddEndPuncttrue
\mciteSetBstMidEndSepPunct{\mcitedefaultmidpunct}
{\mcitedefaultendpunct}{\mcitedefaultseppunct}\relax
\EndOfBibitem
\bibitem[Ackermann \latin{et~al.}(2024)Ackermann, Arnold, Bra{\ss}, Cardot,
  Green, Heinze, Hill, Lu, Macke, Retegan, \latin{et~al.}
  others]{ackermann2024quanty}
Ackermann,~K.; Arnold,~K.; Bra{\ss},~M.; Cardot,~C.; Green,~R.; Heinze,~S.;
  Hill,~P.; Lu,~Y.; Macke,~S.; Retegan,~M., \latin{et~al.}  {QUANTY}, a quantum
  many-body scripting toolkit. \textbf{2024}, \relax
\mciteBstWouldAddEndPunctfalse
\mciteSetBstMidEndSepPunct{\mcitedefaultmidpunct}
{}{\mcitedefaultseppunct}\relax
\EndOfBibitem
\end{mcitethebibliography}

\begin{suppinfo}


\begin{figure}[h!]
 \centering
  \includegraphics[width=0.5\textwidth]{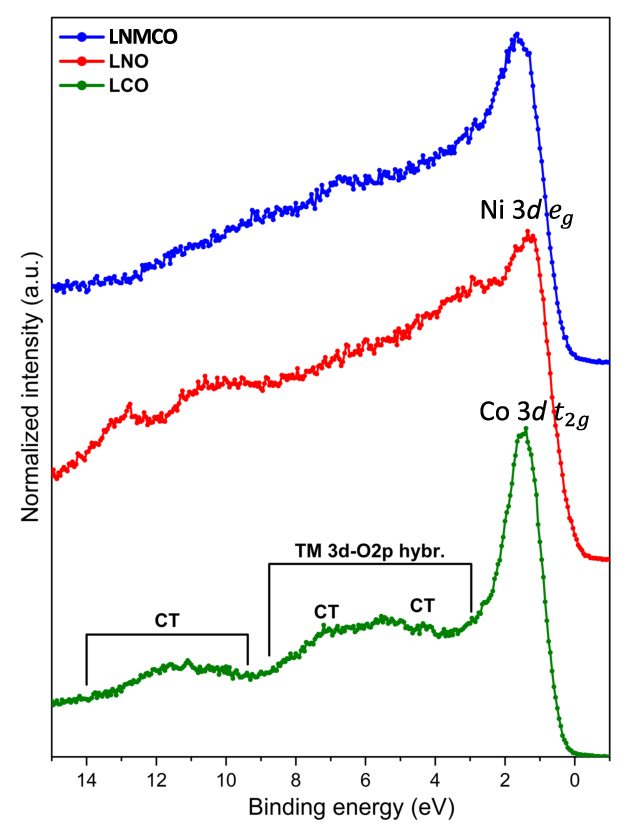}
  \captionsetup{labelformat=empty}
  \caption{Fig. S1: Measured valence band spectra of pristine LCO, LNO and LNMCO.}
\end{figure}

\begin{figure}[h!]
 \centering
  \includegraphics[width=0.5\textwidth]{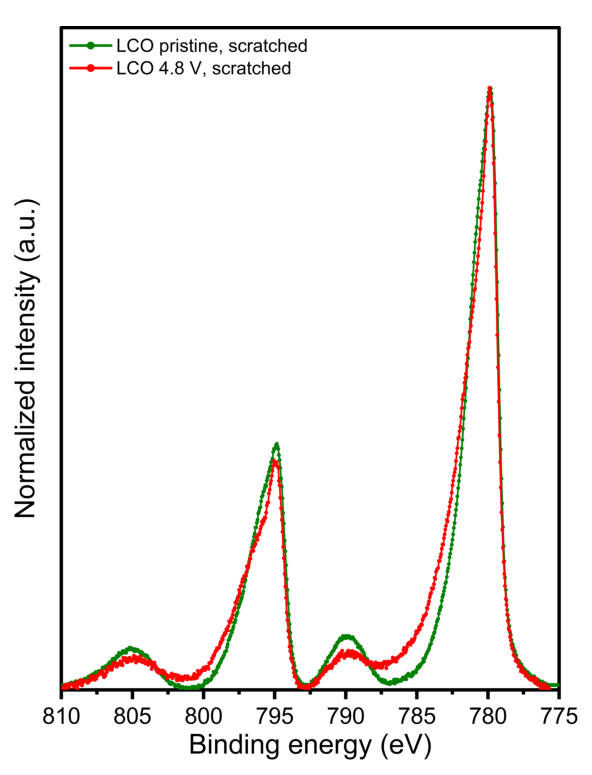}
  \captionsetup{labelformat=empty}
  \caption{Fig. S2: Co 2$p$ spectra of in-vacuo scratched pristine and fully deintercalated LCO.}
  \label{fig:LCO_exp_de}
\end{figure}

\end{suppinfo}


\end{document}